\documentclass{mn2e}
\usepackage{times}
\input{psfig.sty}
\newif\ifAMStwofonts

\title{Radio Emission as a Test of the Existence of Intermediate Mass
Black Holes in Globular Clusters and Dwarf Spheroidal Galaxies}

\author[Maccarone] {Thomas J. Maccarone\\ Astronomical Institute ``Anton
Pannekoek'', University of Amsterdam, Kruislaan 403, 1098 SJ,
Amsterdam, The Netherlands}

\date{}

\begin{document}

\maketitle

\label{firstpage}

\def\simlt{\mathrel{\rlap{\lower 3pt\hbox{$\sim$}}
        \raise 2.0pt\hbox{$<$}}}
\def\simgt{\mathrel{\rlap{\lower 3pt\hbox{$\sim$}}
        \raise 2.0pt\hbox{$>$}}}

\input epsf

\begin{abstract}

We take the established relation between black hole mass, X-ray
luminosity, and radio luminosity and show that intermediate mass black
holes, such as those predicted to exist at the centers of globular
clusters, will be easily identifiable objects in deep radio
observations.  We show that the radio observations will be far more
senstive than any possible X-ray observations. We also discuss the
likely optical photometric and spectroscopic appearance of such
systems in the event that radio detections are made.

\end{abstract}

\begin{keywords}
globular clusters:general -- globular clusters:individual:Omega Cen -- accretion, accretion discs -- black hole physics -- radio continuum:general
\end{keywords}

\section{Introduction}

Intermediate mass black holes may represent the link between the
stellar mass black holes seen in the Milky Way (which are probably all
less massive than 20$M_\odot$ - see McClintock \& Remillard 2003) and
the supermassive black holes thought to exist in the centers of
galaxies (which are mostly more massive than $10^6 M_\odot$, but see
e.g. Fillipenko \& Ho 2003 for a galaxy with a slightly less massive
nucleus).  It is possible that the AGN have grown from $\sim 300
M_\odot$ black holes formed from core collapses of Population III
stars (see e.g. Fryer, Woosley \& Heger 2001).  The centers of dense
star clusters may also produce intermediate mass black holes, either
through the coalescence of massive stars into a single extremely
massive star which then undergoes core collapse into a black hole
(Portegies Zwart \& McMillan 2002; G\"urkan, Freitag \& Rasio 2004) or
through mergers of smaller black holes (Miller \& Hamilton 2002).
Searches for these black holes have, to date, proved inconclusive.

Evidence for the existence of black holes in binary systems in the
Milky Way has been based largely on measuring the masses of X-ray
binaries from their rotation curves along with estimates of their
inclination angles and donor star masses (see Orosz 2003 for a
review).  When the measured mass is larger than 3$M_\odot$, the object
is assumed to be a black hole.  In galactic nuclei, measurement of the
mass of the central dark object has come from a handful of techniques.
For bright active galactic nuclei (AGN), reverberation mapping, which
uses the light travel times to regions with a particular ionization
level (see e.g. Peterson 1998 for a review), can be applied.  For
galaxies with weak or undetected AGN activity, other techniques are
required.  One such method is measurement of the rotational velocity
and the velocity dispersion of the stars or gas in a galaxy's disk;
gas measurements are capable of probing smaller scales, while stellar
measurements are far less likely to be affected by non-gravitational
pertubrations and hence are more reliable (see Kormendy \& Richstone
1995 for a review).  Furthermore, rotation curves are measured on
distance scales larger than that where the black hole's mass dominates
a galaxy's total enclosed mass.

A popular alternative is to use only the velocity dispersion of the
galaxies' bulge stars (Ferrarese \& Merritt 2000; Gebhardt et
al. 2000b), a quantity which correlates well with the rotational
velocity curve and can be easier to measure.  Finally, it has recently
been found that the light profile of the inner regions of elliptical
galaxies and of the bulges of spiral galaxies deviates slightly from
the de Vaucouleurs profile, and that these deviations correlate well
with the central black hole mass measured using other techniques
(Graham et al. 2001).  The results of Graham et al. (2001) are not
well understood on theoretical grounds.

All these techniques are rather difficult to apply to black holes in
globular clusters.  While rotation measurements for globular clusters
are possible (see e.g. Gebhardt et al. 2000a; Anderson \& King 2003),
they are rather difficult to obtain and have not yet been sufficient
for proving the existence of a central black hole.  More recent
attempts to measure the dark masses at the centers of Local Group
globular clusters have focused on measuring which regions of
velocity-position phase space are occupied by stars and comparing
these observations with model predictions (Gerssen et al. 2002).

Even M15, which shows the best evidence for rotation has a velocity
dispersion comparable to its rotational velocity in its center
(Gebhardt et al. 2000a; Gerssen et al. 2002). Given the large velocity
dispersion of the stars in globular clusters, and the relatively small
number of stars, even with perfect measurements of the velocities of a
very large sample of the stars, it might not always be possible to
measure the gravitationational potential well enough to measure the
mass of a central dark object (see e.g. the discussion in Drukier \&
Bailyn 2003).  Attempts have been made to use large kinematic data
sets to place constraints on the masses of central compact objects,
but the uncertainties are quite large and the results are highly model
dependent (e.g. van der Marel 2001; Gerssen et al. 2002,2003).
Without reliable measurements to calibrate a relation, use of the
Ferrarese \& Merritt technique or of the Graham et al. technique would
be quite dangerous.

A few techniques have also been suggested that apply only to globular
clusters.  One such method is to search for high velocity stars within
the cluster, but the predictions for how many high velocity stars
should exist given a certain black hole mass depend on an extremely
detailed treatment of the stellar dynamics (Drukier \& Bailyn 2003).
Another possibility is the use of exotic ejected binary systems as
evidence for the existence of the black hole-black hole binary system
in the core.  In NGC 6752, the binary pulsar system PSR J1911-5958A is
seen 3.3 half mass radii from the core of the globular cluster
(D'Amico et al. 2002).  The rather long period circular orbit suggests
that the ejection was caused by a recoil of a binary system
substantially larger than the pulsar binary, but with a large binding
energy.  A natural possibility for such a binary is one containing two
black holes (Colpi, Possenti \& Gualandris 2002).  The cluster also
has a high central mass-to-light ratio (D'Amico et al. 2002).  The
evidence for a black hole-black hole binary at the core of this
cluster is highly circumstantial, though, and should be confirmed by
some other techniques before the hypothesis is accepted.

The use of stellar kinematics to identify intermediate mass black
holes in globular clusters poses challenges not found either in binary
systems or in galaxy centers.  It would thus be wise to consider
additional techniques, both to improve the capability of discovering
these objects, and to bolster confidence in detections made using
stellar kinematics.  Recent work on correlations between the X-ray and
radio properties of both X-ray binaries and AGN indicates that the
ratio of radio to X-ray power increases with black hole mass and
decreases with accretion rate (Gallo, Fender \& Pooley 2003; Merloni,
Heinz \& Di Matteo 2003; Falcke, K\"ording \& Markoff 2003; Maccarone,
Gallo \& Fender 2003).  Given that the central black holes in globular
clusters are likely to be substantially more massive than stellar mass
black holes, and are certainly accreting at low fractions of the
Eddington luminosity (since none have yet been detected in the
X-rays), these results indicate that radio measurements may hold the
key for detecting the central black holes in globular clusters.  In
this Letter, we outline the predictions made for radio flux from
central black holes in globular clusters given certain constraints on
the X-ray luminosity.

\section{The correlations}

Two important results have been found recently that indicate that
radio, rather than X-rays may be the most efficient waveband for
identifying accreting intermediate mass black holes, especially in
Milky Way globular clusters.  The first is the result of Gallo, Fender
\& Pooley (2003), which shows, for stellar mass black holes, that
while the X-ray and radio luminosities of low/hard state objects (see
Nowak 1995 for a discussion of spectral state and Fender 2004 for a
review of the radio properties of X-ray binaries) are tightly
correlated, the radio luminosity falls off more slowly at low
luminosity than does the X-ray luminosity.  Thus, at lower
luminosities, the radio emission becomes relatively easier to detect
compared with the X-ray emission.  The second key recent discovery is
that the radio to X-ray flux ratio increases with the mass of the
accreting black hole (e.g. Heinz \& Sunyaev 2003; Merloni, Heinz \& Di
Matteo 2003).  This indicates that for intermediate mass black holes,
the radio emission will be substantially brighter than for stellar
mass black holes at the same luminosity, making them again easier to
detect.

The correlations do break down at high fractions of the Eddington
luminosity, both in X-ray binaries which accrete from a companion
star, and in AGN, which like black holes at the centers of globular
clusters, are most likely to accrete from the interstellar medium.
High/soft states, where the X-ray spectrum is dominated by a component
well modeled by a geometrically thin, optically thick accretion disk
(Shakura \& Sunyaev 1973) typically set in above about 2\% of the
Eddington luminosity, although some scatter is likely due to
hysteresis effects in state transition luminosities (Maccarone 2003;
Maccarone \& Coppi 2003).  In the high/soft states of X-ray binaries,
the radio emission is highly suppressed (Tananbaum et al. 1972; Fender
et al. 1999), probably because geometrically thin disks have weaker
poloidal magnetic fields than do geometrically thick accretion flows,
and such magnetic fields are required for jets to be launched (Livio,
Ogilvie \& Pringle 1999; Meier 2001).  The radio emission turns back
on at higher fractions of the Eddington rate.  The results of Fender
et al. (1999) for X-ray binaries have recently been extended to AGN,
which also show suppressed jet activity between about 2 and 20\% of
the Eddington accretion rate (Maccarone, Gallo \& Fender 2003).
Nonetheless, any IMBH within the Milky Way's globular clusters should
be in the low/hard state; one in a higher luminosity state would be
quite bright and would have already been known based on its X-ray
properties unless it were highly absorbed.

\section{X-ray constraints for globular clusters}

Recent theoretical work has suggested that all globular clusters
should have central black holes with masses of order $10^{-3}$ times
the stellar mass in the cluster, due to mergers of black holes (Miller
\& Hamilton 2002).  Additionally, young dense stellar clusters are
also expected to produce black holes with $10^{-3}$ times the stellar
mass in the cluster through stellar mergers leading to the production
of a single highly massive star which evolves quickly into a black
hole (Portegies Zwart \& McMillan 2002; G\"urkan et al. 2004).  It is
expected, however, that the young stellar clusters dense enough to
have these stellar mergers in their cores will evolve in a different
way than the globular clusters we see today.  These systems are thus
not relevant to discussions specific to globular clusters, but the
basic point presented here, that intermediate mass black holes at low
luminosities will be more easily detected in radio than in X-ray is as
relevant for these black holes as it is for those that might found in
globular clusters.

Given the theoretical predictions for the masses of black holes in
globular clusters and the proof from pulsar measurements that certain
globular clusters have substantial amount of gas (i.e. about 0.1
atoms/cm$^3$ at a temperature of about 10$^4$ K - see e.g. Freire et
al. 2001), one can compute a Bondi accretion rate.  Given the expected
radiative efficiency at low accretion rate, the expected X-ray
luminosity will be quite low (detectable in some clusters with the
typical {\it Chandra} exposure times, but faint enough that
spectroscopy will be difficult and it will be virtually impossible to
distinguish one of these sources from the other types of less exotic
X-ray sources found in globular clusters based solely on the position
and X-ray properties of the source).

A couple of the Milky Way's globular clusters (47 Tuc and M15) show
enough millisecond pulsars that the gas content can be estimated by
variations in the dispersion measure through the cluster; since these
measurements are consistent with what it expected from stellar mass
loss, it seems reasonable to assume that the similar values are
present in other clusters (Freire et al. 2001).  Recent {\it Chandra}
observations of M15 have shown that there is no ``nucleus'' of the
cluster seen in X-rays to a limit of 5.6$\times10^{32}$ ergs/sec, a
factor of a few thousand below the expected Bondi accretion rate
assuming a radiative efficiency of 0.1$c^2$(Ho, Terashima \& Okajima
2003).  This factor is roughly in line with the radiative efficiency
corrections expected from either an advection dominated accretion flow
(Narayan \& Yi 1994) or a jet dominated accretion flow (Fender, Gallo
\& Jonker 2003).  Additionally, theoretical work more detailed than
Bondi \& Hyole (1944) suggested that the Bondi accretion rate may
represent an overestimate of how much accretion actually takes place.
Such claims are supported by the lack of detections of isolated
neutron stars accreting from the ISM and the very low luminosities of
the central black holes in many elliptical galaxies (Perna et
al. 2003; Di Matteo et al. 2000).  Furthermore, Ho et al. (2003)
assumed a black hole mass of 2000 $M_\odot$, while 0.1\% of the
cluster mass for M15 is about 440 $M_\odot$, which would reduce the
Bondi rate by an additional factor of about 20.  An upper limit has
also been placed on the X-ray luminosity of the core 47 Tuc at
$10^{31}$ ergs/sec, which gives similar constraints on the radiative
efficiency, assuming that the black hole mass is a few hundred solar
masses (Grindlay et al. 2001).

\section {Predictions for the radio flux} 

From the relation of Merloni et al. (2003), we find that
\begin{equation}
F_{5 GHz} = 10 \left(\frac{L_x}{3\times10^{31} {\rm ergs/sec}}\right)^{0.6} 
\left(\frac{M_{BH}}{100 M_\odot}\right)^{0.78} \left(\frac{d}{10 {\rm kpc}}\right)^{-2} {\rm {\mu}Jy}
\label{merloni}
\end{equation}
A slightly higher radio flux can be derived from the relation of
Maccarone, Gallo \& Fender (2003) which removes the objects thought to
be in the high/soft state.

The sensitivities of the deepest existing radio fields (none of which
are of globular clusters) are about 8 ${\mu}$Jy rms, so detections can
be made above about 25${\mu}$Jy.  For the controversial case of M15,
where Gebhardt et al. (2000a) claimed a 2500 $M_\odot$ black hole, and
which is located about 10 kpc from the Sun (Harris 1996), an X-ray
luminosity at the upper limit of the {\it Chandra} data would yield a
radio flux of about 0.7 mJy.  Radio observations with 8 ${\mu}$Jy rms
would be able to place the same constraints on the whether a 2500
$M_\odot$ black hole exists as X-ray measurements a factor of about
250 deeper in flux than the current ones.  Those {\it Chandra}
observations already required about 30 ksec of time. A single 12 hour
observation with the VLA will then place a stronger constraint on
whether there exists a $\sim 1000 M_\odot$ black hole in M15 than
would spending the entire mission lifetime of Chandra on the same
field of view.

Once a strong candidate is observed, there should be a few tests
performed to ensure that the radio source is not a pulsar.  The first,
obviously, is to look for pulsations.  The second would be to measure
the radio spectrum - jets from low/hard state black holes should have
flat radio spectra, while pulsars have quite steep radio spectra.  A
final test is that pulsars scintillate quite strongly, while the jet
from a low luminosity intermediate mass black hole should be far too
large to scintillate (see Narayan 1992 for a discussion of
scintillation).  Additionally, pulsars are strongly polarized (see
e.g. Manchester, Han \& Qiao 1998), while low/hard state radio spectra
are polarized at only the 1-2\% level (Fender 2001).

Similar reasoning can be applied to the Milky Way's dwarf spheroidal
galaxies with only a few small changes.  These systems are even harder
to work with than the globular clusters for typical measurement
techniques, since the density of stars in the center is much lower.
Furthermore, the theoretical expectations for the masses of central
black holes have not been worked out in as much detail.  Finally, only
the Sagittarius dwarf is near enough that it is likely to be
detectable with this technique.

\section{Prediction for the optical/infrared appearance}

Low/hard state X-ray binaries and from luminosity AGN show a
flat-to-slightly inverted radio spectrum (i.e. $\alpha\simlt0$, where
$F_\nu\propto$$\nu^{-\alpha}$), presumably due to synchrotron
self-absorption of a more intrinsically steep power law, in agreement
with the conical jet model (Blandford \& K\"onigl 1979).  The cutoff
frequency where the jet becomes optically thin is found to be in the
mid-to-near-infrared for jets from X-ray binaries (Corbel \& Fender
2002), and should be at a somewhat longer wavelength for a globular
cluster black hole which is weak in X-rays, since the cutoff
frequency, $\nu_\tau$, scales as
$M_{BH}^{1/3}(\frac{\dot{m}}{\dot{m}_{EDD}})^{2/3}$ for a radiatively
inefficient accreting black hole with a typical spectral index
(i.e. $\alpha=0.5$) for the optically thin component of the jet
emission (see equation 14 of Heinz \& Sunyaev 2003).  One should then
expect that the optical emission from the jet will be optically thin,
and have a spectral index of about $\alpha=0.5$.  Given a cutoff at 10
$\mu$m for the globular cluster black holes' jets, one finds that:

\begin{equation}
V= 23.0 - 2.5 {\rm log}\left(\frac{F_{5 GHZ}}{10 {\mu}{\rm{Jy}}}\right),
\end{equation}
or combining with equation \ref{merloni}
\begin{equation}
V= 23.0 - 1.5 {\rm log}\left(\frac{L_x}{3\times10^{31} {\rm ergs/sec}}\right) -1.95  
{\rm log}\left(\frac{M_{BH}}{100 M_\odot}\right) + 5 {\rm log} \left(\frac{d}{10 {\rm kpc}}\right)
\end{equation}
and the optical colors should be approximately $V-I$=0.3 and $V-K$=1.2
(all assuming no reddening).  Given the density of stars in the core
of a globular cluster and the difficulty in performing accurate
photometry there, optical photometry is a less robust technique for
proving that a black hole is present than radio detection.
Furthermore, there may be additional non-jet flux in this range of
wavelengths due to emission from a pre-shock jet region (e.g. Markoff,
Falcke \& Fender 2003) or due to optical synchrotron emission from a
magnetically or otherwise powered corona above the accretion disk (see
e.g. Merloni, Di Matteo \& Fabian 2000).  A perhaps more robust proof
would come from spectroscopy; detection of a faint red star which
shows no evidence for any spectral lines (since the continuum should
be dominated by non-thermal emission) would make non-accretion models
for the optical emission rather difficult.  It is unlikely that the
thermal component from the geometrically thin part of the accretion
disk will contribute substantially to the flux; the thin disk
components of low/hard state black holes begin to emit a substantial
fraction of the total luminosity only in the ultraviolet (see
McClintock et al. 2001), and in any event, it is not clear whether the
accreted interstellar matter will have enough angular momentum even to
form a disc at large radii.  Spectroscopic studies of faint stars in
the cores of globular are rather difficult to perform, so the
spectroscopic method is likely to be of use only to confirm the
results of radio and photometric work, and not as the starting point
for the search for black holes in globular clusters.

\section{Which GCs are the best candidates?}
We make crude estimates of the expected radio fluxes, under the
following assumptions: the black hole mass is 0.1\% of the globular
cluster mass, the actual accretion rate is 0.1\% of the Bondi rate
(the value considered most likely in Perna et al. 2003), a gas density
of 0.15 H cm$^{-3}$ (approximately the value measured for both 47 Tuc
and M15) and the radiative efficiency scales as shown in Fender, Gallo
\& Jonker (2003; FGJ). We note, also, that if there is a binary black
hole in the center of the globular cluster, the Bondi rate may be
considerably reduced due to orbital motions.  This will be a problem
only if the mass ratio between the two black holes is less than about
10 this problem or if the central binary is quite hard.  Miller \&
Hamilton (2002) find that if black hole binaries exist in globular
clusters, they are likely to have large mass ratios and to spend most
of their lifetimes in wide separations, so the problem of the binary
orbit is unlikely to be a serious one.

It is suggested in FGJ that the accretion power goes into two
components - a relativistic jet and X-ray emission, and it is shown
that the jet power will scale as $A$$L_X^{0.5}$, where $A$ is a
constant found to be at least $6\times10^{-3}$ for black holes and
$L_X$ is the X-ray luminosity.  The total luminosity, $L_{tot}$ is
then equal to $A$$L_X^{0.5}+L_X.$ Assuming that the efficiency in
converting accreted mass into energy is the standard 10\%, then the
radiative efficiency in the X-rays will be:
\begin{equation}
\eta=(0.1)\times\left(1+\frac{A^2}{2 L_{tot}}-A\sqrt{\frac{A^2}{4 L_{tot}^2}+\frac{1}{L_{tot}}}\right),
\end{equation}
and we set $A$ to the most conservative value for jet power of
$6\times10^{-3}$, and where $L_{tot}$ is expressed in Eddington units.
We note that the radiative efficiency scales with the accretion rate
in quite a similar way in the JDAF and ADAF models.

The Bondi rate (Bondi \& Hoyle 1944) is given in convenient form by Ho
et al. (2003):
\begin{equation}
\dot{M}_{BH} = 3.2\times10^{17} \left(\frac{M_{BH}}{2000 M_\odot}\right)^2
\left(\frac{n}{0.2 {\rm cm^{-3}}}\right) \left(\frac{T}{10^4 {\rm K}}\right)^{3/2} {\rm g
s}^{-1}.
\end{equation}
We make two sets of calculations, assuming the accretion rate to be
$10^{-2}$ and $10^{-3}$ of the Bondi rate (the proposed maximum
accretion rate and most likely accretion rate from Perna et al. 2003).
We note that for the higher accretion rate, the predictions for the
X-ray luminosity and the radio flux from 47 Tuc are in excess of the
observed upper limits, so unless the black hole mass in 47 Tuc is
anomalously low, the values in the table for the $10^{-2}$ Bondi
accretion rate case are too optimistic.  It also seems likely that a
2.4 mJy source in Omega Cen would have been missed, but there are not
any good upper limits published, and Omega Cen is sufficiently far
south to have been missed by surveys such as NVSS.

We take the data for the Milky Way's globular clusters from Harris
(1996).  The mass is not reported directly, so we convert from
integrated absolute visual magnitude $M_V$ to mass using a
mass-to-$V$-light ratio of 1.8 (Piatek et al. 1994).  The
globular cluster's mass is then:
\begin{equation}
M_{GC}=10^6 M_\odot 10^{-(M_V+10)/2.5},
\end{equation}
and the central black hole mass is predicted, from Miller \& Hamilton
(2002), to be:
\begin{equation}
M_{BH}=1000 M_\odot 10^{-(M_V+10)/2.5}.
\end{equation}

\begin{table*}
\begin{tabular}{lllllllll}
\hline
 GC&               $M_{BH}$ & $d$ &$L_{X,-3}$ &  $F_{5GHz,-3}$ & $F_{5GHz,-2}$ &      Common name & comments\\
\hline
 NGC 5139      &   1250 & 5.1  &1.1$\times10^{31}$ & 154&  2400&      Omega Cen\\
 NGC 104       &   560 & 4.3  &1.0$\times10^{30}$ & 27&   430&    47 Tuc&$L_X<10^{31}$ from obs.\\
 NGC 6656      &   240& 3.2  &8.1$\times10^{28}$ & 6 & 90&    M22\\
 NGC 6388      &   810 & 11.5 & 3.1$\times10^{30}$ & 10&160\\
 NGC 6266      &   450 & 6.7  &5.4$\times10^{29}$ & 7 & 100&    M62\\
 NGC 2808      &   550 & 9.3  &9.8$\times10^{29}$ & 6& 90\\
 NGC 6441      &   470 & 9.7  &6.1$\times10^{29}$ & 3& 50&LMXB\\
 NGC 6273      &   410 & 8.5  &4.0$\times10^{29}$ & 3& 50&     M19\\
 NGC 6402      &   390 & 8.7  &3.4$\times10^{29}$ & 3& 40&     M14\\
 NGC 5904      &   320 & 7.3  &1.9$\times10^{29}$ & 2& 40&     M5\\
 NGC 7078      &   440 & 10.2 & 5.0$\times10^{29}$ & 3&  40&   M15&$L_X<5\times10^{32}$ from obs.\\
 NGC 6440      &   300 & 8.  &1.6$\times10^{29}$ & 2& 30&LMXB\\
 NGC 6715      &   960 & 26.2 & 5.3$\times10^{30}$ & 3& 50&   M54&Sag dwarf - BH mass may be higher\\
 NGC 6626      &   210 & 5.7 & 5.1$\times10^{28}$ & 1& 20&  &   M28\\
 NGC 6205      &   250 & 7.0 & 9.3$\times10^{28}$ & 1&20\\
\end{tabular}
\caption{The globular clusters expected to be brightest in radio.  The
black hole mass is expressed in solar units, the distance in
kiloparsecs, the X-ray luminosity in ergs/sec and the radio flux in
$\mu$Jy; raw data is taken from Harris (1996) and the computation of
the parameters is described in the text.  $F_{5GHz,-3}$ refers to the
prediction assuming the accretion rate is 0.1\% of Bondi and
$F_{5GHz,-2}$ refers to the prediction assuming 1\% of Bondi. The
X-ray luminosities assuming the 1\% of Bondi accretion rate are all
about 100 times the luminosities assuming 0.1\% of Bondi accretion
rates, since the X-ray efficiencies are all in the $\dot{m}^2$
regime.}
\label{predict_table}
\end{table*}

The predicted radio and X-ray fluxes for 15 good candidate globular
clusters are presented in table \ref{predict_table}, along with the
estimates of the black hole mass and distance used.  We include some
objects calculated to lie below the detection limit from a reasonable
integration largely because the estimate of the fraction of the Bondi
accretion rate that is actually accreted computed in Perna et
al. (2003) is highly uncertain, and it is well within the realm of
possibility that the expected fluxes could be at least a factor of 10
higher.  The strongest observational point in favor of the claim by
Perna et al. (2003) is that no isolated neutron stars accreting from
the interstellar medium were detected by ROSAT; in fact there are
numerous unidentified sources in the ROSAT catalogs, and some fraction
could be these isolated neutron stars. 

We note that M54 in particular might be a substantially better
candidate than the table suggests - it is the nucleus of the
Sagittarius dwarf spheroidal galaxy, and hence setting its black hole
mass value to $10^{-3}$ times the globular cluster's mass, rather than
$10^{-3}$ times the galaxy's mass may be a substantial underestimate.
Since, in our model, the radio luminosity scales roughly as $M^2$, an
underestimate of the black hole mass by a factor of 10 has profound
implications on the expected radio flux.  We also note that NGC 6440
and NGC 6441 have bright low mass X-ray binaries (Liu, van Paradijs \&
van den Heuvel 2001), which could seriously affect the accuracy of
X-ray flux measurements of their nuclei.  Furthermore, the expanded
VLA (see http://www.aoc.nrao.edu/evla/index.shtml) is expected to have
a sensitivity limit a factor of five times lower than the current VLA,
and should have a higher angular resolution to ensure that
extragalactic confusion will still not be a problem.

The two clusters with the highest predicted flux levels are far enough
south that they cannot be observed with the VLA.  The observations
which come closest to the predicted flux limits of any of these
clusters are the ATCA observations of 47 Tuc (Fruchter \& Goss 2000;
McConnell \& Ables 2000).  These reach an rms flux density of only
about 30 $\mu$Jy, and are taken at $\approx$ 20 cm wavelengths where
the angular resolution is poor, and pulsars are relatively brighter
further reducing their usefulness for finding a black hole's radio
emission.  We find no reports in the literature regarding deep radio
observations of Omega Cen, either; the ATCA archives contain only
10-hour observations of the outskirts of the cluster, and these
observations are in taken in a multi-channel spectral line mode with
low bandwidth that makes the data not sufficiently deep for detecting
continuum at the levels predicted.  It seems possible that the
brightest clusters may have radio emission from central black holes
that is detectable given a deep, optimally designed observation, but
which could not be detected with existing data.

\section{Conclusions}

Determining whether globular clusters contain intermediate mass black
holes is a key problem in astronomy.  In this Letter, we have outlined
a technique for using radio and optical photometry to test whether
there are good candidates for intermediate mass black hole in the
centers of globular clusters.  Furthermore, this technique is truly
sensitive to the mass of the central compact object, unlike most
kinematic methods which are merely sensitive to dark matter
concentrations.  We have also shown that the results are likely to
apply to dwarf spheroidal galaxies.  Most importantly, we have shown
that there exists a strong possibility for detecting at least a
handful of the putative central black holes of Milky Way globular
clusters through their radio emission in the near future, and have
shown that radio emission is likely to be the most effective way to
identify the strongest candidates for having central black holes.
Obviously there are considerable uncertainties regarding the predicted
level of radio emission, but given a detection in the radio and either
a detection or a strong upper limit in the X-rays, it may be possible
to provide good evidence for the existence of an intermediate mass
black hole in a globular cluster or a dwarf spheroidal galaxy.

\section{Acknowledgments}
I wish to thank Russell Edwards, Rob Fender, Elena Gallo, Andrea
Merloni, Cole Miller, Simon Portegies Zwart, Fred Rasio and Ben
Stappers for useful communications.  I also thank the anonymous
referee for comments which have greatly improved the clarity of the
article, and which have also improved the content of the article,
especially in the discussion of stellar dynamics.

\label{lastpage}

\begin{thebibliography}{}
\bibitem{}Anderson, J., King, I.R., 2003, AJ, 126, 772
\bibitem{}Blandford, R.D., K\"onigl, A., 1979, ApJ, 232, 34
\bibitem{}Bondi, H., Hoyle, F., 1944, MNRAS, 104, 273
\bibitem{}Colpi, M., Possenti, A., Gualandris, A., 2002, ApJL, 570, 85
\bibitem{}Corbel, S., Fender, R.P., 2002, ApJL, 573, 35
\bibitem{}D'Amico, N., Possenti, A., Fici, L., Manchester, R.N., Lyne, A.G., Camile, F., Sarkissian, J., 2002, ApJL, 570, 89
\bibitem{}Di Matteo, T., Quataert, E., Allen, S.W., Narayan, R., Fabian, A.C., 2000, MNRAS, 311, 507
\bibitem{}Drukier, G.A., Bailyn, C.D., 2003, ApJL, in press (astro-ph/0309514)
\bibitem{}Falcke, H., K\"ording, E. \& Markoff, S., 2003, A\&A Let., submitted (FKM03)
\bibitem{}Fender, R.P. et al., 1999, ApJL, 519, 165
\bibitem{}Fender, R.P., 2001, MNRAS, 322, 31
\bibitem{}Fender, R.P., Gallo, E. \& Jonker, P., 2003, MNRAS, 343L, 99
\bibitem{}Fender, R.P., 2004, in {\it Compact Stellar X-ray Sources}, Cambridge University Press (eds. W.H.G. Lewin \& M. van der Klis), astro-ph/0303339
\bibitem{}Ferrarese, L. \& Merritt, D., 2000, ApJL, 539, 9
\bibitem{}Fillipenko, A.V., Ho, L.C., 2003, ApJL, 588, 13
\bibitem{}Freire, P.C., Kramer, M., Lyne, A.G., Camilo, F.,
Manchester, R.N. \& D'Amico, N., 2001,ApJL, 557, 105
\bibitem{}Fruchter, A.S., Goss, W.M., 2000, ApJ, 536, 865
\bibitem{}Fryer, C.L., Woosley, S.E., Heger, A., 2001, ApJ, 550, 372
\bibitem{}Gallo, E., Fender, R.P. \& Pooley, G.G., 2003, MNRAS, 344, 60
\bibitem{}Gebhardt, K., Pryor, C., O'Connell, R.D., Williams, T.B. \& Hesser,J.E., 2000a, AJ, 119, 1268
\bibitem{}Gebhardt, K. et al., 2000b, ApJL, 539, 13
\bibitem{}Gerssen, J., van der Marel, R.P., Gebhardt, K., Guhathakutra, P., Peterson, R.C., Pryor, C., 2002, AJ, 124, 3270
\bibitem{}Gerssen, J., van der Marel, R.P., Gebhardt, K., Guhathakutra, P., Peterson, R.C., Pryor, C., 2003, AJ, 125, 376
\bibitem{}Graham, A.W., Erwin, P., Caon, N., Trujillo, I., 2001, ApJL, 563, 11
\bibitem{}Grindlay, J.E., Heinke, C., Edmonds, R.D., Murray, S.S., 2001, Science, 292, 2290
\bibitem{}G\"urkan, M.A., Freitag, M., Rasio, F.A., 2004, ApJ,
submitted (astro-ph/0308449)
\bibitem{}Harris, W.E., 1996, AJ, 112, 1487
\bibitem{}Heinz, S., Sunyaev, R.A., 2003, MNRAS, 343L, 59 
\bibitem{}Ho, L., 2002, ApJ, 564, 120
\bibitem{}Ho, L.C., Terashima, Y., Okajima, T., 2003, ApJL, 587, 35
\bibitem{}Kormendy, J., Richstone, D., 1995, ARA\&A, 33, 581
\bibitem{}Liu, Q.Z., van Paradijs, J., van den Heuvel, E.P.J., 2001, A\&A, 368, 1021
\bibitem{}Livio, M., Ogilvie, G.I., Pringle, J.E., 1999, ApJ, 512, 100
\bibitem{}Maccarone, T.J., 2003, A\&A, 409, 697
\bibitem{}Maccarone, T.J., Coppi, P.S., 2003, MNRAS, 338, 189
\bibitem{}Maccarone, T.J., Gallo, E., Fender, R.P., 2003, MNRAS, 345, L19
\bibitem{}Manchester, R.N., Han, J.L., Qiao, G.J., 1998, MNRAS, 295, 280
\bibitem{}Markoff, S., Falcke, H. \& Fender, R.P., 2001, A\&A Let., 372, 25
\bibitem{}McClintock, J.E. et al., 2001, ApJ, 555, 477 
\bibitem{}McClintock, J.E. \& Remillard, R.A., 2003, to appear in {\it
Comapct Stellar X-ray Sources} (astro-ph/0306213)
\bibitem{b24}Meier, D.L., 2001, ApJL, 548, 9L
\bibitem{}Meier, D.L., Koide, S. \& Uchida, Y., 2001, Science, 291, 84
\bibitem{}Merloni, A., Di Matteo, T. \& Fabian, A.C., 2000, MNRAS, 318, L15
\bibitem{}Merloni, A., Heinz, S. \& Di Matteo, T., 2003, MNRAS, 345, 1057
\bibitem{}Miller, M.C., Hamilton, D.P., 2002, MNRAS, 330, 232
\bibitem{}Narayan, R., 1992, Phil. Trans. Roy. Soc. Lon. A, 341, 151
\bibitem{b27}Narayan, R. \& Yi, I., 1994, ApJL, 428, 13L
\bibitem{}Nowak, M.A., 1995, PASP, 107, 1207
\bibitem{}Orosz, J.A., 2003, IAUS, 212, 365
\bibitem{}Perna, R., Narayan, R., Rybicki, G., Stella, L., Treves, A., 2003, ApJ, 594, 936
\bibitem{}Peterson, B.M., 1998, AdSpR, 21, 57
\bibitem{}Piatek, S., Pryor, C., McClure, R.D., Fletcher, J.M., Hesser, J.E., 194, AJ, 107, 1397
\bibitem{}Portegies Zwart, S.F., McMillan, S.L.W., 2002, ApJ, 576, 899
\bibitem{b31}Shakura, N.I. \& Sunyaev, R.A., 1973, A\&A,  24, 337
\bibitem{b34}Tananbaum, H., Gursky, H., Kellogg, E., Giacconi, R. \& Jones, C.,1972, ApJL, 177, 5L
\bibitem{}van der Marel, R., 2001, in Black Holes in Binaries and Galactic Nuclei, 246
\end{thebibliography}
\end{document}